# Minimization of Handoff Failure Probability for Next-Generation Wireless Systems

Debabrata Sarddar[1], Tapas Jana[2], Souvik Kumar Saha[1], Joydeep Banerjee[1] , Utpal Biswas[3], M.K. Naskar[1]

1. Department of Electronics and Telecommunication Engg, Jadavpur University, Kolkata – 700032. E-mail:dsarddar@rediffmail.com, jogs.1989@rediffmail.com, souviksaha@ymail.com, mrinalnaskar@yahoo.co.in .

2. Department of Electronics and Communication Engg, Netaji Subhash Engg College, Techno City, Garia, Kolkata – 700152. Email: tjanansec@gmail.com,

3. Department of Computer Science and Engg, University of Kalyani, Nadia, West Bengal, Pin- 741235, Email: utpal01in@yahoo.com

***ABSTRACT***

***During the past few years, advances in mobile communication theory have enabled the development and deployment of different wireless technologies, complementary to each other. Hence, their integration can realize a unified wireless system that has the best features of the individual networks. Next-Generation Wireless Systems (NGWS) integrate different wireless systems, each of which is optimized for some specific services and coverage area to provide ubiquitous communications to the mobile users. In this paper, we propose to enhance the handoff performance of mobile IP in wireless IP networks by reducing the false handoff probability in the NGWS handoff management protocol. Based on the information of false handoff probability, we analyze its effect on mobile speed and handoff signaling delay.***

## 1. INTRODUCTION

A cell is the radio area covered by a transmitting station or a Base Station (BS). All Mobile Terminals (MTs) within that area are connected and serviced by the BS. Therefore, ideally, the area covered by a cell is a circle, with the BS being at the centre. Thus, actually cells are not hexagonal. Hexagon fitted the planed area nicely and hexagon is the greatest area in the circle with respect to any other shape. The cell is therefore approximated to a regular hexagon and side of the hexagon is the common cord of two adjacent cells. When any MT crosses a common cord of a cell, we can say that handoff has occurred from one cell to another cell.

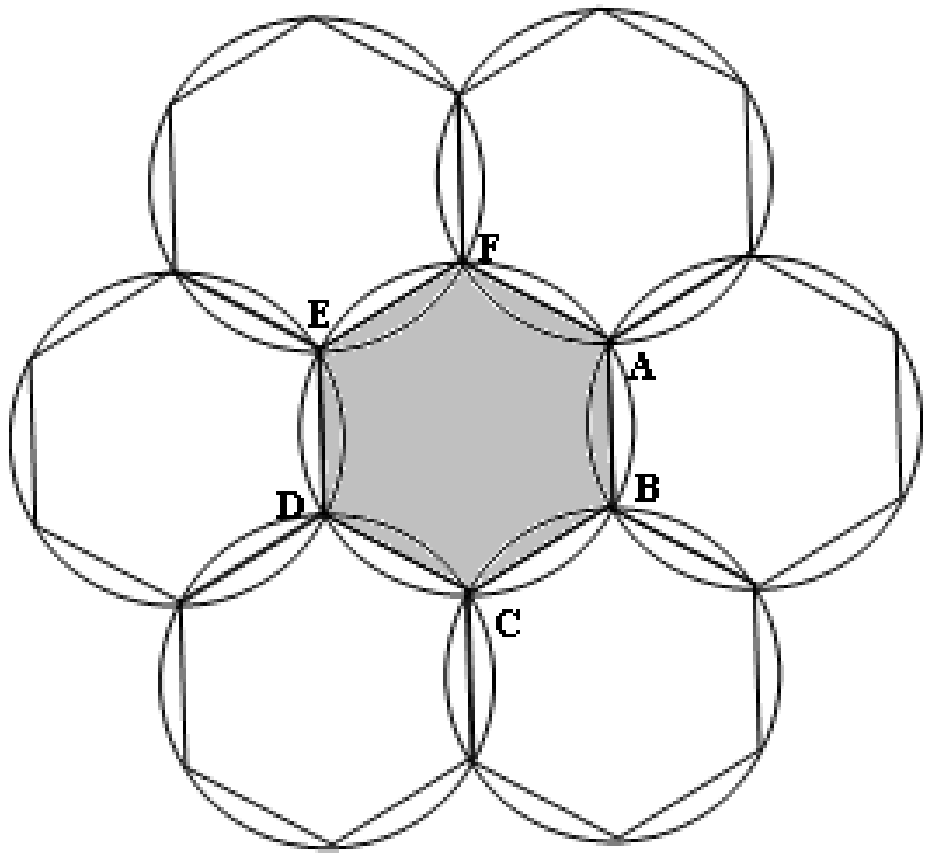

Fig. [1]. Seven Cell cluster.

We consider a seven cell cluster in Fig. [1]. ABCDEF is a regular hexagon. Here, these adjacent seven cells are fitted in such a manner that no vacant places are left between them and the common cord between two circular cells is the side of the hexagon. As the hexagon fitted the planed area nicely so we can imagine that the shape of the cell is a regular hexagon and it is the maximum area in the circular cell (shaded area) with respect to any other shape.

Today's wireless world provides several communication networks, such as Bluetooth for personal area, IEEE 802, for local area, UMTS (Universal Mobile Telecommunications System) for wide area and satellite networks for global networking. These networks are complementary to each other. The best feature of the individual networks is to provide ubiquitous 'always best connection' [13] to the mobile users [14].

Mobility management contains two components: location management and handoff management [5]. Location management helps to track the locations of mobile users between consecutive communications. But handoff management process keeps its connection active even when it moves from one base station (BS) to another. Location management techniques for NGWS [8],[16] can be used in Architecture for ubiquitous Mobile Communications (AMC). But seamless support of handoff management in NGWS is an open issue [17].

In real scenario the integrated architecture may consists of many different wireless systems. In NGWS, two types of handoffs arise: horizontal handoff and vertical handoff [11] .

- Horizontal handoff: handoff between two BSs of the same system. It can be further classified into
- 1) Link-layer handoff: Horizontal handoff between two BSs, under the same Foreign Agent (FA), *e.g.*, the handoff of a MT from BS10 to BS11 in Fig. [2].
- 2) Intra-system handoff: Horizontal handoff between two BSs that belong to two different FAs and both FAs belongs to the same system and hence to same gateway foreign agent (GFA), *e.g.*, handoff of MT from BS11 to BS12 in Fig. [2].
- Vertical handoff (Inter-System Handoff): Handoff between two BSs, belong to two different systems and two different GFAS, *e.g.*, the handoff of the MT from BS12 to BS20 in Fig. [2].

In this paper, we do not address the link-layer handoff. Our work will be on intra-system and inter-system handoff [17]. The large value of signaling delay associated with the intra-system and inter-system handoff [18] can be detrimental for delay-sensitive real-time services. We therefore try to minimize the signaling delay by reducing the probability of false handoff.

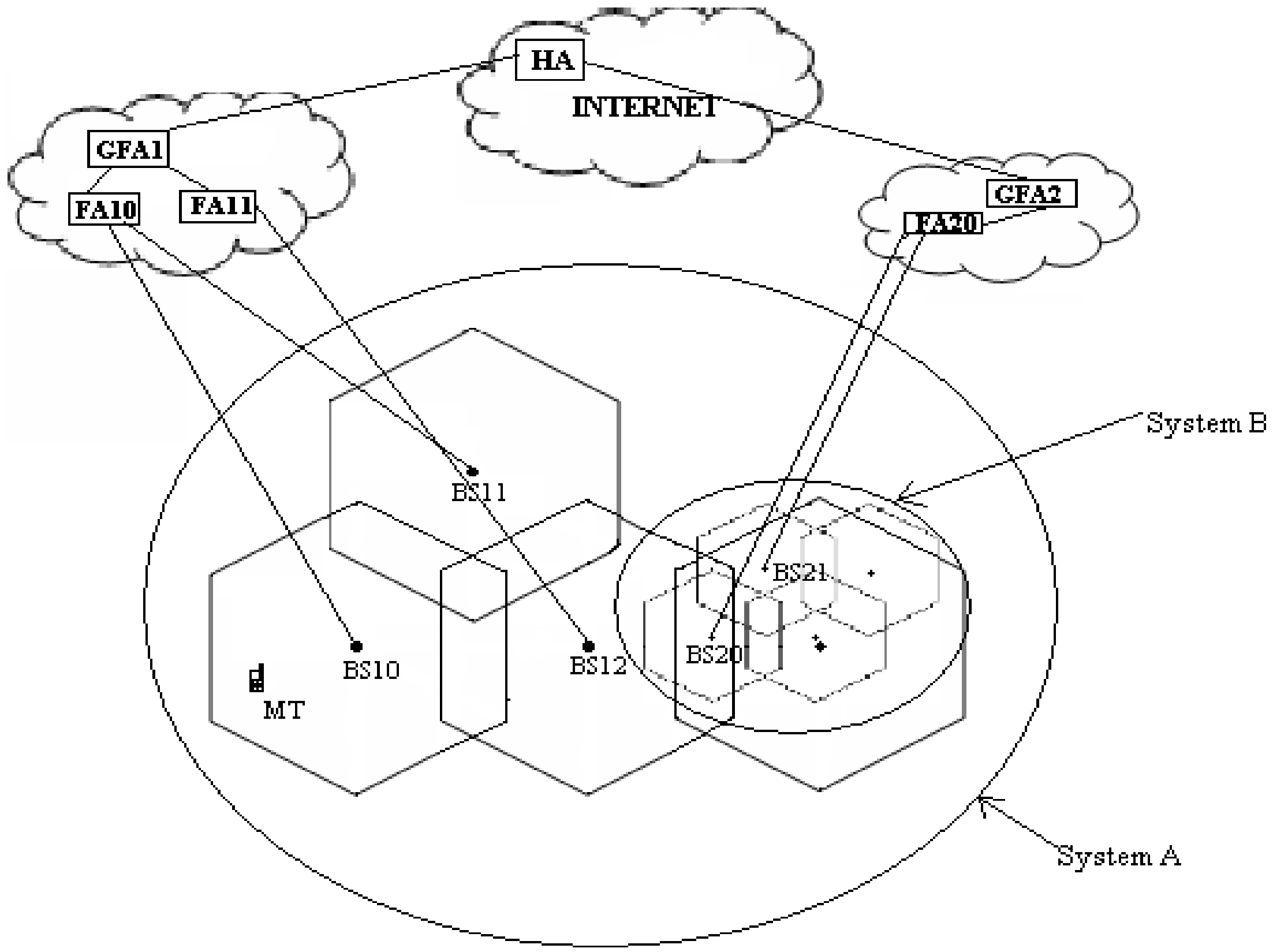

Fig. [2] Handoff in the integrated NGWS architecture

HA: Home Agent
GFA: Gateway Foreign Agent
FA: Foreign Agent
MT: Mobile Terminal
BS: Base Station

## 2. RELATED WORKS

Handoff management protocols operating from different Layers of the classical protocol stack (e.g., link layer, network layer, transport layer, and application layer) have been proposed in the literature [17]. [1] focuses on integrating layer 3 handoff messages into layer 2 messages. It reduces handoff latency employing FMIP-based WiBro system. In [2], we are introduced to a new enhanced Handoff Protocol for Integrated Networks (eHPINs), which localizes the mobility management enabling fast handoff. Application layer mobility using Session Initiation Protocol (SIP) is proposed in [18]. SIP based mobility does not require any changes to the IP stack of the mobile users.

Hierarchical Mobile IP [20] and other micro-mobility protocols such as cellular IP [21], IDPM [22], and HAWAII [23] address the problem of high global signaling load and handoff latency by introducing another layer of hierarchy to the MIP architecture to localize the signaling messages to one domain. MIP and micro-mobility solutions [23, 22, and 21] achieve reduction in registration signaling delay, but fail to address the problem of handoff requirement detection delay [17].

[3] decreases the number of control packets for proactive caching and also proposes a superior replacement caching algorithm; which together enables reduction in handoff delay. A generic link layer technique is used in [9] to add the handoff protocols operating from the upper layers. Different link layer assisted handoff algorithms to enhance Received Signal Strength (RSS) value and thus reduce the handoff latency and handoff failure are proposed in [23] and [26].

We assume from the link-layer assisted handoff protocols implicitly that the handoff latency of the intra-system and inter-system handoff are constant. Based on this protocol, the link-layer assisted handoff protocols initiate the handoff when the RSS of the serving BS goes below a pre-defined fixed threshold value. In fact, signaling delay [11] of the intra-system and inter-system handoff depends on the traffic level in the backbone network, the wireless link quality and distance between the user and its home network at the handoff instance. So, a fixed delay for intra-system and inter-system handoff has poor performance when the handoff signaling delay varies.

## 3. PROPOSED METHOD

Previously, we have discussed why the shape of the cell is considered to be a regular hexagon. We have seen two cells overlapping in such a manner that the common cord between two adjacent circular cells also becomes the common side of the regular hexagons, when the cells are considered to be hexagons. However, two cells may overlap in such a way that there is some overlapping hexagonal portion between them. Such type of structure is given below in Fig. [3].

Here in Fig. [3], AB is the side of regular hexagonal cell served by the Old BS (OBS). But $A'B'$ is the common cord of the two adjoining cells, one served by the OBS and the other by the NBS. When a Mobile Terminal (MT) crosses $A'B'$ , then it will be under the New BS (NBS). This is because RSS of NBS is greater then RSS of OBS to the right side of $A'B'$ .

Once the MT reaches the boundary of the circular cell (real) then the MT discovers that it may enter into the coverage area. Here, we have considered some hexagonal portion to be overlapping. MT is moving from its current serving BS (old BS), to the future serving BS (new BS). This is shown in details in Fig. [3]., from which the probability of false handoff is calculated. The following are the definitions of the notations used in the figure:

- Sth : The RSS threshold value to initiate the handoff. This implies, when RSS of OBS goes below Sth, the Hierarchical Mobile IP (HMIP) registration procedures are initiated for MT's handover to the NBS.
- Smin: The MT's minimum RSS value to communicate successfully between an MT and BS.
- OBS: The old BS.
- NBS: The new BS.
- Here a: The cell size served by a BS (i.e., the length of each side of hexagonal cells).
- P: The point when the MT's RSS from the OBS drops below Sth. On the point 'P' the MT understands that it is on the overlapping position.
- Here d: The distance from the hexagonal cell boundary to the point 'P'.
- Here θ: The motion direction of MT from point 'P' to handoff to NBS.
- L: Distance between side of hexagon and common cord of two hexagons.
- β: Any direction of MT when at time 't' it covered the distance 'x' from the point 'P'.

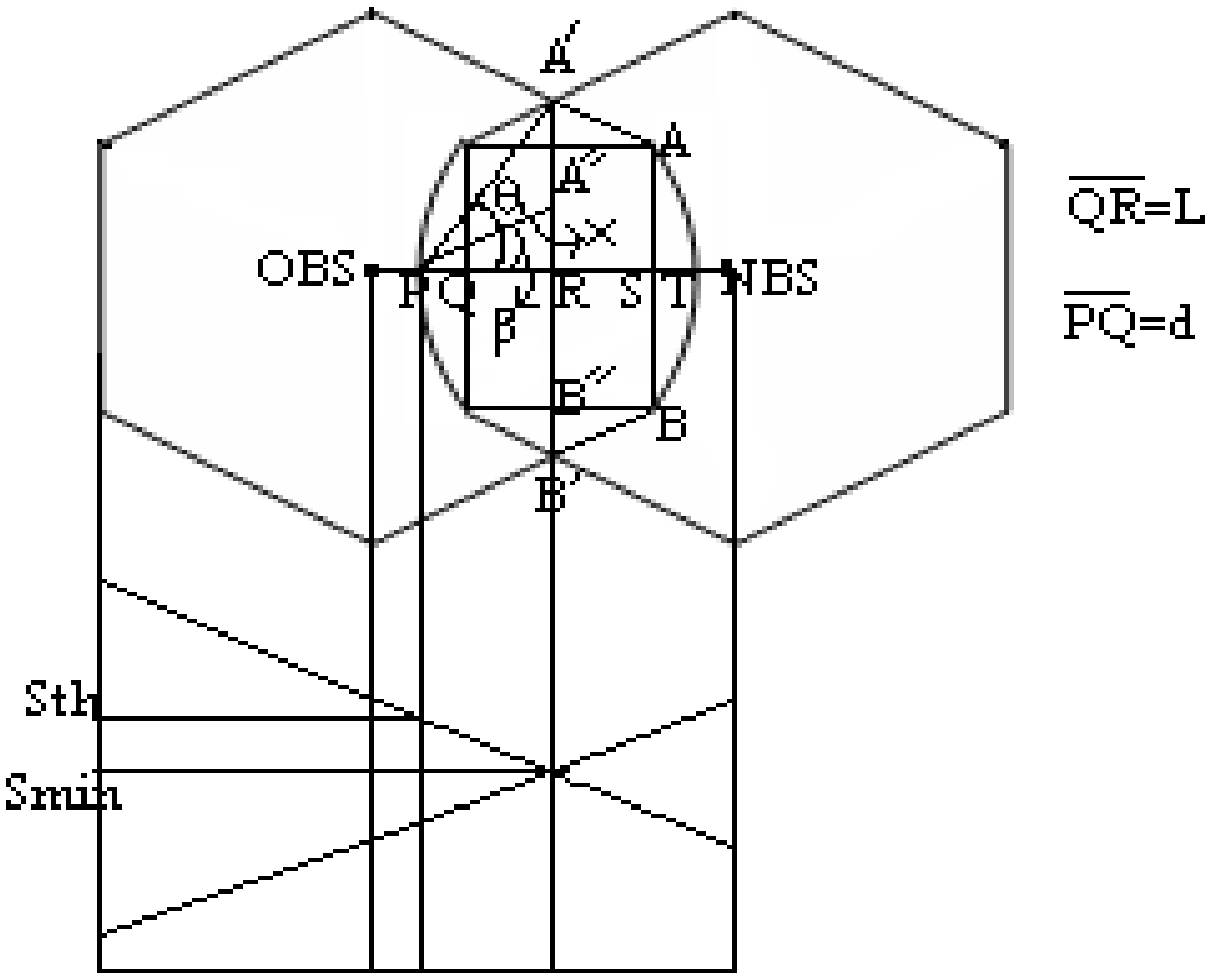

Fig. [3]. Analysis of the handoff process.

From the figure, we get:-

AB = a = radius of the circle = length of the side of proposed hexagon.

OQ = √3a ⁄ 2

PQ = d = OP – OQ

= a – (√3a ⁄ 2)

= (2a - √3a) / 2

QR = L (assumption)

When MT crosses the line A′B′, only then handoff will occur.

PR = PQ + QR

= d + L

= (2a - √3a + 2L) / 2

A′A″ = L tan30º

= L / √3

A′ R = (√3a + 2L) / 2√3

X = ((2a - √3a + 2L) sec β) / 2

t = ((2a - √3a + 2L) sec β) / 2 v

Where ‘v’ is the velocity of MT.

$\tan\theta_1$ = A′ R / PR

= (√3a + 2L) /√3 (2a - √3a + 2L)

A scenario where an MT is currently served by OBS is considered for the analysis. We consider that the MT is moving with a speed ‘V’. ‘V’ is assumed to be uniformly distributed in [Vmin, Vmax]. So we can say that the probability density function (pdf) of ‘V’ is given by

$$f_v(v) = \frac{1}{V_{\max} - V_{\min}} \quad ; \quad V_{\max} > V > V_{\min} . \quad \ldots\ldots\ldots\ldots\ldots\ldots\ldots (1)$$

During the course of movement the MT discovers that it is going to move into the subnet served by NBS. It is assumed that during the course of its movement when the MT reaches the point P, the RSS from OBS goes below Sth. So, when MT reaches P, the HMIP registration is initiated with the New FA (NFA). At this point, the RSS received by the MT from NBS may not be sufficient for the MT to send the HMIP registration messages to NFA through NBS. Hence, the MT may send the HMIP registration message to NFA through OBS. This is called pre-registration. For a smooth and successful handoff from OBS to NBS, MT's HMIP registration with NFA and link layer associations with NBS must be completed before the RSS of OBS goes below $S_{min,}$ i.e, before the MT moves beyond the coverage area of OBS.

When the MT is located at point P, it is assumed that it can move in any direction with equal probability, i.e., the probability density function of MT's direction θ is

$$f_\theta(\theta) = \frac{1}{2\pi}; \pi > \theta > -\pi \;\ldots\ldots\ldots\ldots\ldots\ldots\ldots (2)$$

It is clear that the need for handoff to NBS arises only if MT's direction of motion from P is in the range [ $\theta \in (-\theta_1, \theta_1)$ ] where

$\theta_1 = \tan^{-1}[\frac{\sqrt{3}a + 2L}{\sqrt{3}(2a - \sqrt{3}a + 2L)}]$, otherwise the handoff initiation is a false one. Therefore using (2), the probability of false handoff initiation is

$$P_a = 1 - \int_{-\theta_1}^{\theta_1} f_\theta(\theta)d\theta$$

$$= 1 - \frac{\theta_1}{\pi}$$

$$= 1 - \frac{1}{\pi}\tan^{-1}[\frac{\sqrt{3}a + 2L}{\sqrt{3}(2a - \sqrt{3}a + 2L)}] \;\ldots\ldots\ldots\ldots\ldots (3)$$

So we can say false handoff initiation is independent of d but it is dependent on L.
If we consider L = 0, then

$$P_a = 1 - \frac{1}{\pi} \times \frac{5\pi}{12}$$

$$= 1 - \frac{5}{12}$$

$$= \frac{7}{12}$$

= constant.

When the direction of motion of the MT from P, $\beta \in [(-\theta_1, \theta_1)]$, the time it takes to move out of the RSS of OBS is given by

$$t = \frac{\left(2a + \sqrt{3}a + 2L\right)\sec\beta}{2V}. \quad \ldots\ldots\ldots\ldots\ldots\ldots\ldots (4)$$

We know that the pdf of $\beta$ is given by

$$f_\beta(\beta) = \begin{Bmatrix} \frac{1}{2\theta_1}; where -\theta_1 < \beta < \theta_1 \\ 0; otherwise \end{Bmatrix} \ldots\ldots\ldots\ldots\ldots\ldots\ldots (5)$$

Where from (4) 't'is a function of $\beta, i.e., t = g(\beta)$,

Where $g(\beta) = \frac{\left(2a + \sqrt{3}a + 2L\right)\sec\beta}{2V}$,

Therefore, the pdf of t is given by

$$f_t(t) = \sum \frac{f_\beta(\beta_i)}{|g'(\beta_i)|}, \quad \ldots\ldots\ldots\ldots\ldots\ldots\ldots (6)$$

Where $\beta_i$ are the roots of the equation $t = g(\beta) in [-\theta_1, \theta_1]$. The equation $t = g(\beta)$ has two roots in the interval $[-\theta_1, \theta_1]$ and, for each of these roots, $f_\beta(\beta_i) = \frac{1}{2\theta_1}$,

for i = 1 and 2. Therefore, (6) becomes

$$f_t(t) = \frac{1}{\theta_1 |g'(\beta_i)|}, \quad \ldots\ldots\ldots\ldots\ldots\ldots\ldots (7)$$

Where $g'(\beta)$ is the derivative of $g(\beta)$ given by

$$\begin{aligned} g'(\beta) &= \frac{\left(2a + \sqrt{3}a + 2L\right)\sec\beta\tan\beta}{2V} \\ &= t\tan\beta \\ &= t\sqrt{\left(\sec^2\beta - 1\right)} \\ &= t\sqrt{\left[\left(\frac{2Vt}{2a + \sqrt{3}a + 2L}\right)^2 - 1\right]}. \quad \ldots\ldots\ldots\ldots\ldots\ldots (8) \end{aligned}$$

From the equation (7) and (8) we have, the pdf of t is given by

$$f_t(t)=\begin{cases}\dfrac{\left(2a+\sqrt{3}a+2L\right)}{\theta_1 t\sqrt{\left[(2Vt)^2-(2a+\sqrt{3}a+2L)^2\right]}};where\dfrac{\left(2a-\sqrt{3}a+2L\right)}{2V}<t<\dfrac{\sqrt{\dfrac{(2a-\sqrt{3}a+2L)^2}{4}+\dfrac{(\sqrt{3}a+2L)^2}{12}}}{V}\\ 0;otherwise\end{cases}.(9)$$

The probability of handoff failure is given by

$$P_f=\begin{cases}1;where\,\tau>\dfrac{\sqrt{\left[\left(\dfrac{2a-\sqrt{3}a+2L}{2}\right)^2+\left(\dfrac{\sqrt{3}a+2L}{2\sqrt{3}}\right)^2\right]}}{V}\\ P(t<\tau);where\dfrac{2a-\sqrt{3}a+2L}{2V}<\tau<\dfrac{\sqrt{\left[\left(\dfrac{2a-\sqrt{3}a+2L}{2}\right)^2+\left(\dfrac{\sqrt{3}a+2L}{2\sqrt{3}}\right)^2\right]}}{V}\\ 0;where\,\tau\le\dfrac{2a-\sqrt{3}a+2L}{2V}\end{cases}...(10)$$

Where $\tau$ is the handoff signaling delay and $P(t<\tau)$ is the probability that $t<\tau$.

When $\dfrac{2a-\sqrt{3}a+2L}{2V}<\tau<\dfrac{\sqrt{\left[\left(\dfrac{2a-\sqrt{3}a+2L}{2}\right)^2+\left(\dfrac{\sqrt{3}a+2L}{2\sqrt{3}}\right)^2\right]}}{V}$

Using (9) we have

$$P(t<\tau)=\int_0^{\tau} f_t(t)dt$$

$$=\int_{\frac{2a-\sqrt{3}a+2L}{2V}}^{\tau}\frac{(2a+\sqrt{3}a+2L)}{\pi t\sqrt{(2Vt)^2-(2a+\sqrt{3}a+2L)^2}}dt$$

$$=\frac{1}{\theta_1}\cos^{-1}\left[\frac{2a-\sqrt{3}a}{2V\tau}\right]........................................................(11)$$

Now, using (10) and (11) we have

$$P_f = \left\{ \begin{array}{l} 1;\ where\ \tau > \dfrac{\sqrt{\left[\left(\dfrac{2a-\sqrt{3}a+2L}{2}\right)^2 + \left(\dfrac{\sqrt{3}a+2L}{2\sqrt{3}}\right)^2\right]}}{V} \\ \dfrac{1}{\theta_1}\cos^{-1}\left[\dfrac{2a-\sqrt{3}a}{2V\tau}\right];\ where\ \dfrac{2a-\sqrt{3}a+2L}{2V} < \tau < \dfrac{\sqrt{\left[\left(\dfrac{2a-\sqrt{3}a+2L}{2}\right)^2 + \left(\dfrac{\sqrt{3}a+2L}{2\sqrt{3}}\right)^2\right]}}{V} \\ 0;\ where\ \tau \le \dfrac{2a-\sqrt{3}a+2L}{2V} \end{array} \right\}.$$

……………………………………………………………………………………… (12)

## 4. PERFORMANCE ANALYSIS OF PROPOSED SCHEME

We can provide a detailed discussion about the performance of HMIP handoff algorithms using the mathematical formulations derived above.

### 4.1. FALSE HANDOFF INITIATION PROBABILITY

From equation (3) we can infer that, if we unnecessarily increase the value of L, the probability of false handoff initiation increases. This results in the wastage of limited wireless system resources. Moreover, this increases the load on the network that arises because of the handoff initiation. The relationship between probability of false handoff initiation and L is shown in Fig. [4] for a different cell size, ‘a’. Fig. [4] shows that, for a particular value of a, the probability of false handoff initiation increases as L increases. We can also see that the problem of false handoff initiation becomes more and more severe when the size of the cell decreases. The capacity of data rate may increase if the cell size of wireless system decreases. However, our primary target is to reduce the probability of false handoff initiation. For this reason, we have to adjust the value of L in such a way that the probability of false handoff initiation will be lowest. In order to get a constant value of the probability of false handoff initiation, we put the value of L=0, for which we get the value to be 7/12.

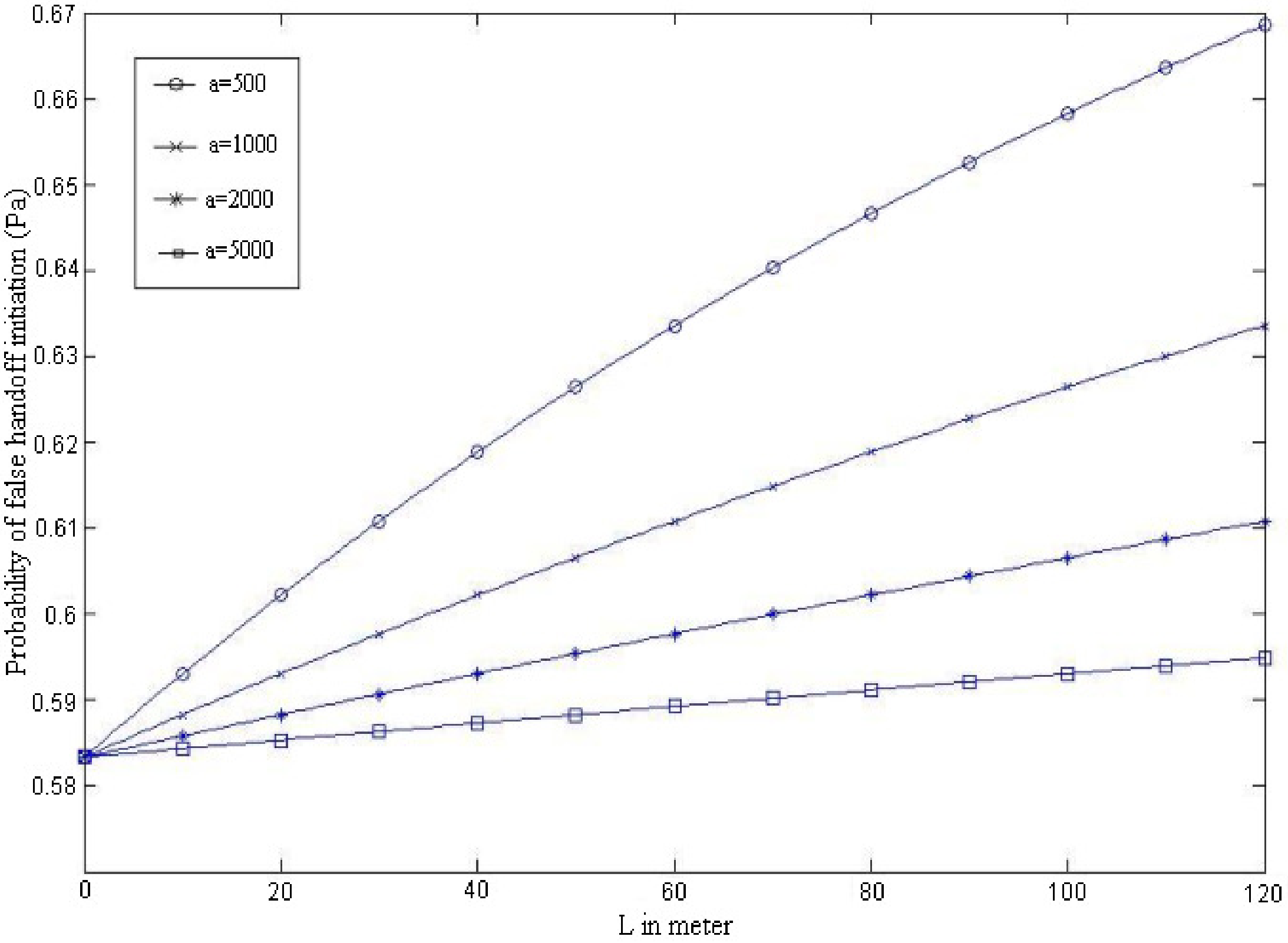

Fig. [4].Relationship between false handoff initiation probability and L.

## 4.2. RELATIONSHIP BETWEEN HANDOFF FAILURE PROBABILITY AND SPEED

From equation (12), we can say that, when $\frac{2a-\sqrt{3}a+2L}{2V} < \tau < \frac{\sqrt{\left[\left(\frac{2a-\sqrt{3}a+2L}{2}\right)^2+\left(\frac{\sqrt{3}a+2L}{2\sqrt{3}}\right)^2\right]}}{V}$, for a fixed value of Sth (and hence a fixed value of corresponding L), handoff failure probability depends on the speed of the MT. In fact, the probability of handoff failure $P_f$ is directly proportional to the speed of MT. If the speed of MT is V, then we can write $P_f \infty V$ . We testify this proportionality with the help of a simulation. For our simulation, we consider that the cell size is 1 km. Fig. [5] (a) and Fig. [5] (b) shows the relationship between the handoff failure probability and MT's speed for intra-system and inter-system handoff, respectively. These figures show the numerical value of $P_f$ for different values of L (corresponding to different values of Sth). The main difference between intra- and inter-system handoff is the latency associated with the handoff process. The latency of inter-system handoff is significantly greater than that of intra-system handoff. This is because during an inter-system handoff, before HMIP registration, authentication and billing procedures are carried out adding extra delay to the handoff process. Moreover, when the inter-system HMIP signaling messages are handed by MT's home agent (HA) instead of gateway foreign agent (GFA), adding extra delay to the signal propagation as the distance of MT from HA is typically larger than that of MT from the GFA. We consider the handoff latency ( $\tau$ ) is 1.5 second for intra-system handoff and 3 second for inter-system handoff. Fig. [5] (a) and Fig. [5] (b) show that for a particular value of L, as speed increases, handoff failure probability increases for both intra- and inter-system handoff. This is because in both the cases, MT requires less time to cross the coverage region of OBS. However, for a particular value of Sth, $P_f$ becomes higher for inter-system handoff compared to intra-system handoff for any speed value.

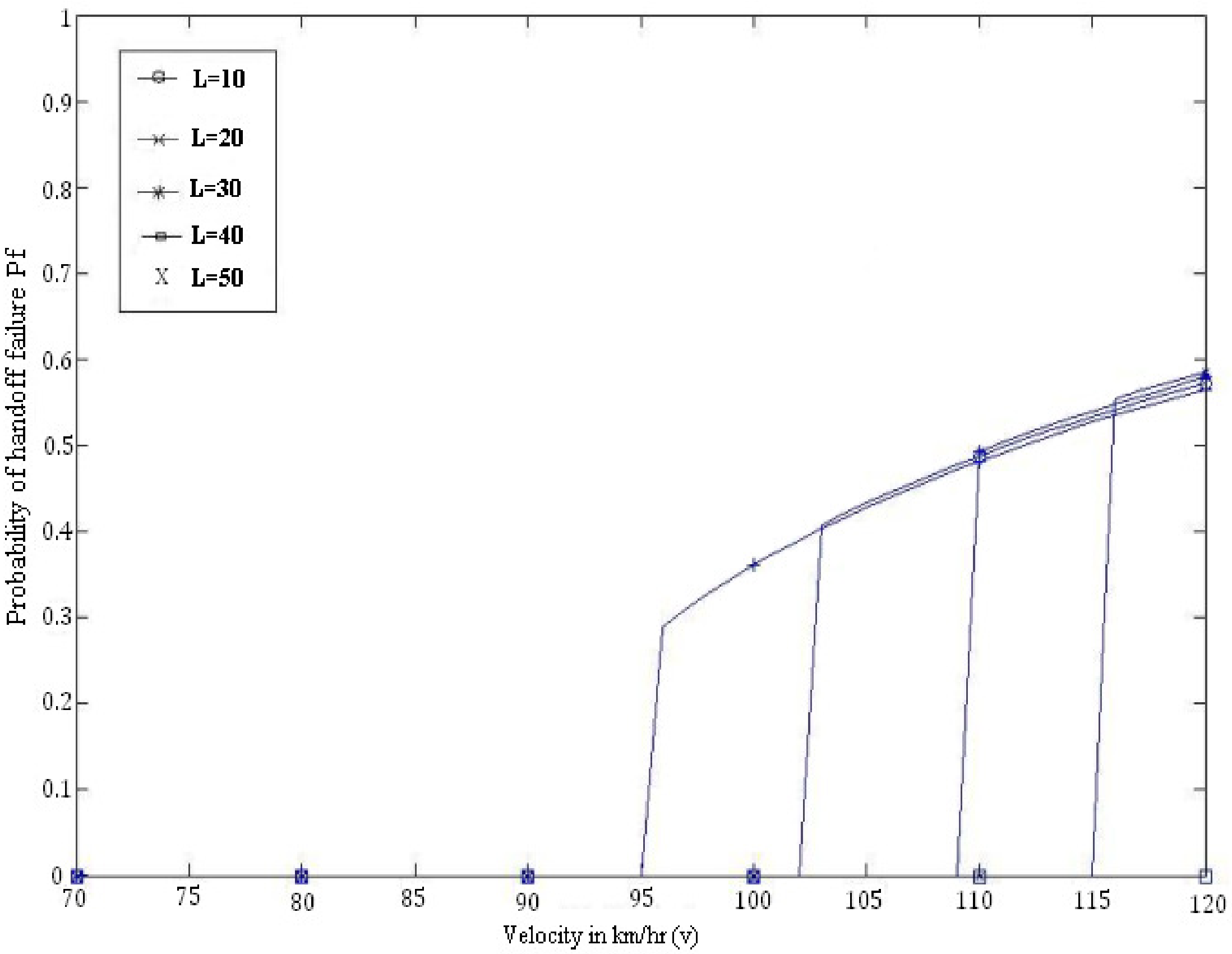

Fig. [5]. (a) Relationship between handoff failure probability and MT's speed for intrasystem handoff with $\tau = 1.5\sec$.

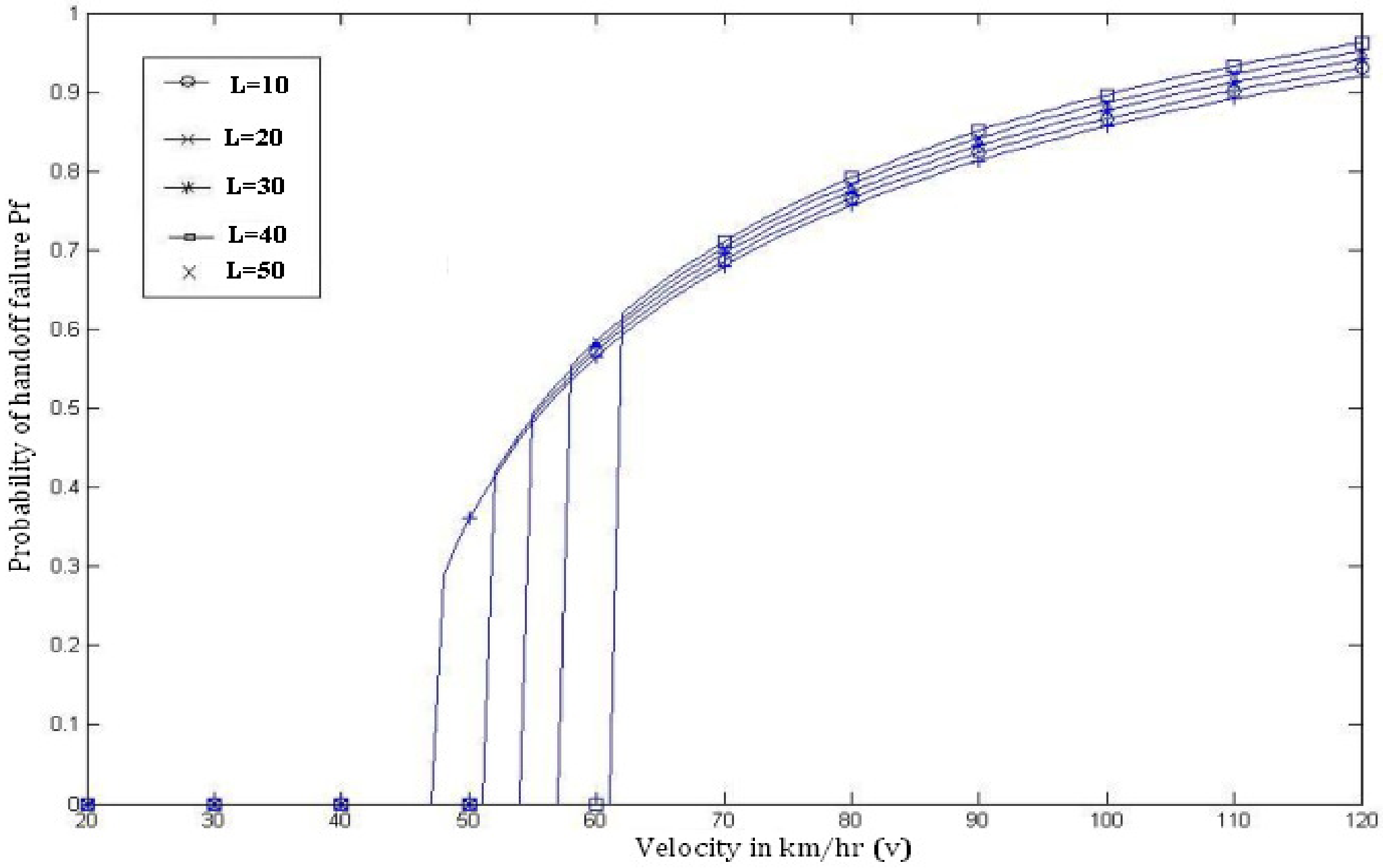

Fig. [5]. (b) Relationship between handoff failure probability and MT's speed for intersystem handoff with $\tau = 3\sec ond.$

## 4.3. RELATIONSHIP BETWEEN HANDOFF FAILURE PROBABILITY AND HANDOFF SIGNALING DELAY

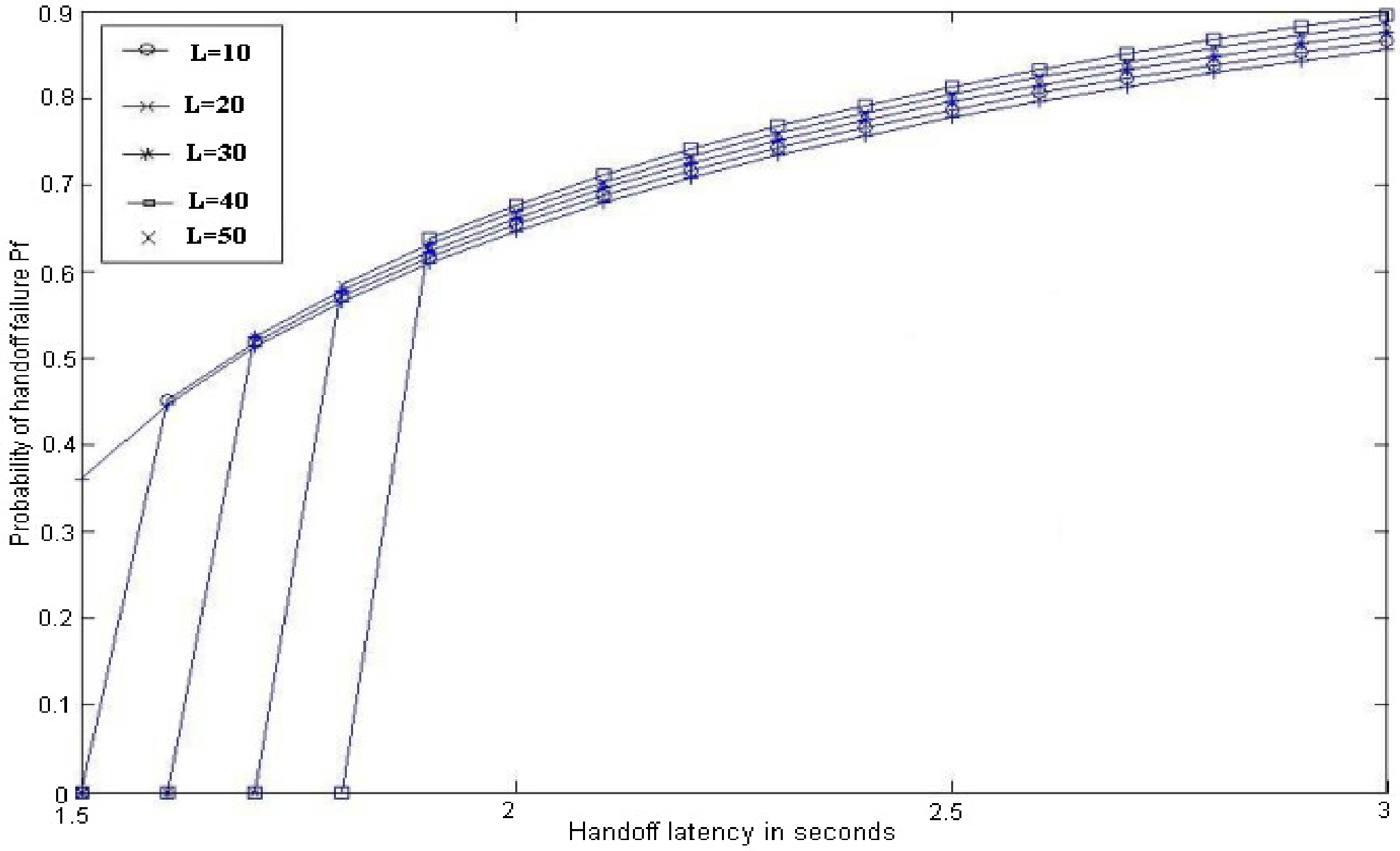

Fig. [6]. Relationship between handoff failure probability and $\tau$.

Fig. [6] shows the relationship between handoff failure probability and handoff signaling delay $(\tau)$ when a fixed value of $S_{th}$ ( and thus, a fixed value of L) is used. We observe that the handoff failure probability increases as handoff signaling delay increases. So, we can say that, in order to keep the handoff failure probability limited, it is essential to predict the handoff signaling delay in advance and accordingly use an adaptive value of Sth. So, we can say that, an unnecessarily large value of L (and hence a large value of Sth) should not be used as it increases the probability of false handoff initiation.

## 5. CONCLUSION, COMPARATIVE STUDY AND FUTURE WORK

In this work, we first discuss the different types of handoff in the next generation wireless systems. Then we analyze the performance of handoff management protocols that use a fixed value of RSS threshold (Sth) to initiate the handoff process. Through our analysis, we observe that when a fixed value of Sth is used, handoff failure probability increases when either speed or handoff signaling delay increases. Based on this analysis, we suggest a method by which handoff failure probability can be kept constant and within limit.

As suggested in [10], handoff initiation increases as d increases. It is also shows that the problem of false handoff initiation becomes more and more severe when the cell size decreases. Actually, at the position P, in Fig. [3], MT can only understand that it is entering in the overlapping area of two adjacent cells. Position P is on the circumference of the cell with the NBS at the centre. Distance from P to hexagonal cell side is ‘d’ [10].

In our problem, the value of d depends on cell size ‘a’ where ‘a’ is the radius of the circular cell and also the length of the hexagon side and the relation between ‘a’ and ‘d’ is given by:-

$$d = \frac{\left(2-\sqrt{3}\right)}{2}a .$$

For overlapping cells, we bring into consideration the variable ‘L’, which denotes the distance between side of hexagon and common cord of the two hexagons. We now find that the false handoff probability no longer depends on the value of d, but on the value of L. When L=0, the value of false handoff initiation is constant and found to be $\frac{7}{12}$ (putting L=0 in the equation (3)) . Also, when L=0, we find that false handoff probability is independent of cell size ‘a’.

Our conclusion is that, if hexagonal overlapping occurs, there will be a non-zero value of L. In that case, false handoff initiation probability will depend on cell size ‘a’, and hence its value will not be a constant. This value will only be constant for L=0, i.e., only when no overlapping occurs between two hexagonal cells.

In our paper, we have conducted all our calculations assuming tacitly that L=0. However, different other propositions may come up when $L\neq0$, when hexagonal overlapping occurs. In that case, $P_a \neq 7/12$. This calls for further investigation. Besides, handoff failure probability increasing with decreasing cell size may also pose a problem which opens up ample scope of future research.

**Author Biographies**

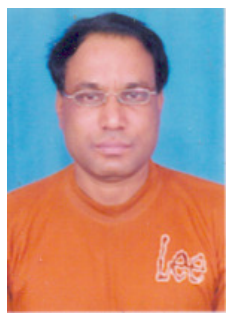

**Debabrata Sarddar** is currently pursuing his PhD at Jadavpur University. He completed his M.Tech in Computer Science & Engineering from DAVV, Indore in 2006, and his B.E in Computer Science & Engineering from Regional Engineering College, Durgapur in 2001. He was earlier a lecturer at Kalyani University. His research interest includes wireless and mobile system.

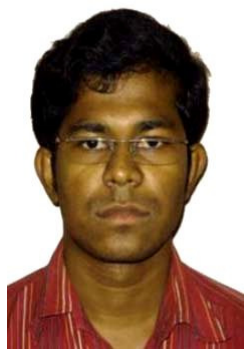

**Tapas Jana** is presently pursuing B.Tech Degree in Electronics and Communication Engg. at Netaji Subhash Engg. College, under West Bengal University Technology. His research interest includes wireless sensor networks and wireless communication systems

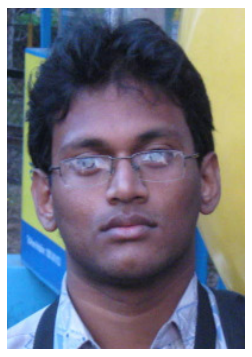

**Souvik Kumar Saha** is presently pursuing B.Tech Degree in Electronics and Telecommunication Engg. at Jadavpur University. His research interest includes wireless sensor networks and wireless communication systems.

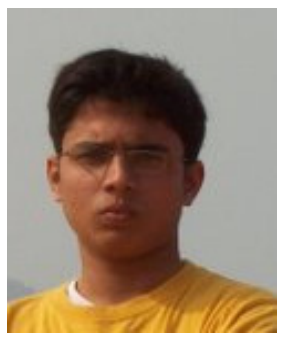

**Joydeep Banerjee** is presently pursuing B.Tech Degree in Electronics and Telecommunication Engg. at Jadavpur University. His research interest includes wireless sensor networks and wireless communication systems.

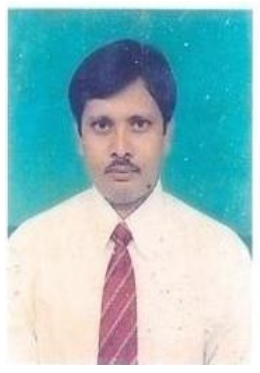

**Utpal Biswas** received his B.E, M.E and PhD degrees in Computer Science and Engineering from Jadavpur University, India in 1993, 2001 and 2008 respectively. He served as a faculty member in NIT, Durgapur, India in the department of Computer Science and Engineering from 1994 to 2001. Currently, he is working as an associate professor in the department of Computer Science and Engineering, University of Kalyani, West Bengal, India. He is a co-author of about 35 research articles in different journals, book chapters and conferences. His research interests include optical communication, ad-hoc and mobile communication, semantic web services, E-governance etc.

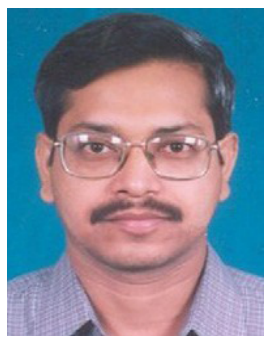

**Mrinal Kanti Naskar** received his B.Tech. (Hons) and M.Tech degrees from E&ECE Department, IIT Kharagpur, India in 1987 and 1989 respectively and Ph.D. from Jadavpur University, India in 2006.. He served as a faculty member in NIT, Jamshedpur and NIT, Durgapur during 1991-1996 and 1996-1999 respectively. Currently, he is a professor in the Department of Electronics and Tele-Communication Engineering, Jadavpur University, Kolkata, India where he is in charge of the Advanced Digital and Embedded Systems Lab. His research interests include ad-hoc networks, optical networks, wireless sensor networks, wireless and mobile networks and embedded systems.

He is an author/co-author of the several published/accepted articles in WDM optical networking field that include "Adaptive Dynamic Wavelength Routing for WDM Optical Networks" [WOCN,2006], "A Heuristic Solution to SADM minimization for Static Traffic Grooming in WDM uni -directional Ring Networks" [Photonic Network Communication, 2006],"Genetic Evolutionary Approach for Static Traffic Grooming to SONET over WDM Optical Networks" [Computer Communication, Elsevier, 2007], and "Genetic Evolutionary Algorithm for Optimal Allocation of Wavelength Converters in WDM Optical Networks" [Photonic Network Communications,2008].